\documentclass[11pt,twoside]{article}
\usepackage{asp2004}
\usepackage{psfig}
\usepackage{epsf}
\usepackage[dvips]{graphicx}
\usepackage{graphics}
\usepackage{lscape}
\def\edcomment#1{\iffalse\marginpar{\raggedright\sl#1\/}\else\relax\fi}
\reversemarginpar

\begin{document}
\title{Type Ia SN1999cw: photometric and spectroscopic study}
\author{F.Bufano$^{1}$, M.Turatto,$^{2}$ S.Benetti$^{2}$,
 A.Harutyunyan$^{1}$, N.Elias de la Rosa$^{2,3}$, E.Cappellaro$^{4}$}
\affil{$^{1}$Universita' di Padova, Italy, $^{2}$Osservatorio Astronomico di Padova, Italy,
  $^{3}$Universidad de La Laguna, Tenerife, Spain, $^{4}$Osservatorio Astronomico di Capodimonte, Napoli, Italy }

\begin{abstract}
The preliminary analysis of the optical data of SN1999cw show that this 
object has the  photometric and  spectroscopic behavior of a Type Ia 
supernova similar to SN1991T, reaching an apparent magnitude at maximum $ B_{max}  = 14.30 $
 and a  $\Delta m_{15}(B)=0.94$. 
\end{abstract}
\thispagestyle{plain}

\section{SN 1999cw}

SN1999cw was discovered on June $28^{th}$,1999 by Johnson \& Li with KAIT telescope, 
$21''.1$ East
 and $1''.5$ South of the nucleus of  MCG-01-02-001 (RA= 0h 20m 1s.46 and Dec= -6$^{\circ}$
 20' 03''.6, J2000), a barred spiral galaxy (SBab), with redshift z= 0.0125 
and galactic extinction $A_{B}^{g}$ = 0.154 mag (Schlegel et al, 1998).
 The preliminary reduction of a  DFOSC spectrum showed that it was a peculiar Type Ia supernova
 (Rizzi et al,1999).
 Immediately after the discovery, we began to observe
 SN1999cw collecting data both in optical and infrared wavelengths.

\section {Photometry and Spectroscopy}
\begin {figure}[!h]
\begin{center}
\includegraphics [ width=4.9cm, height=6.6cm, angle=270]{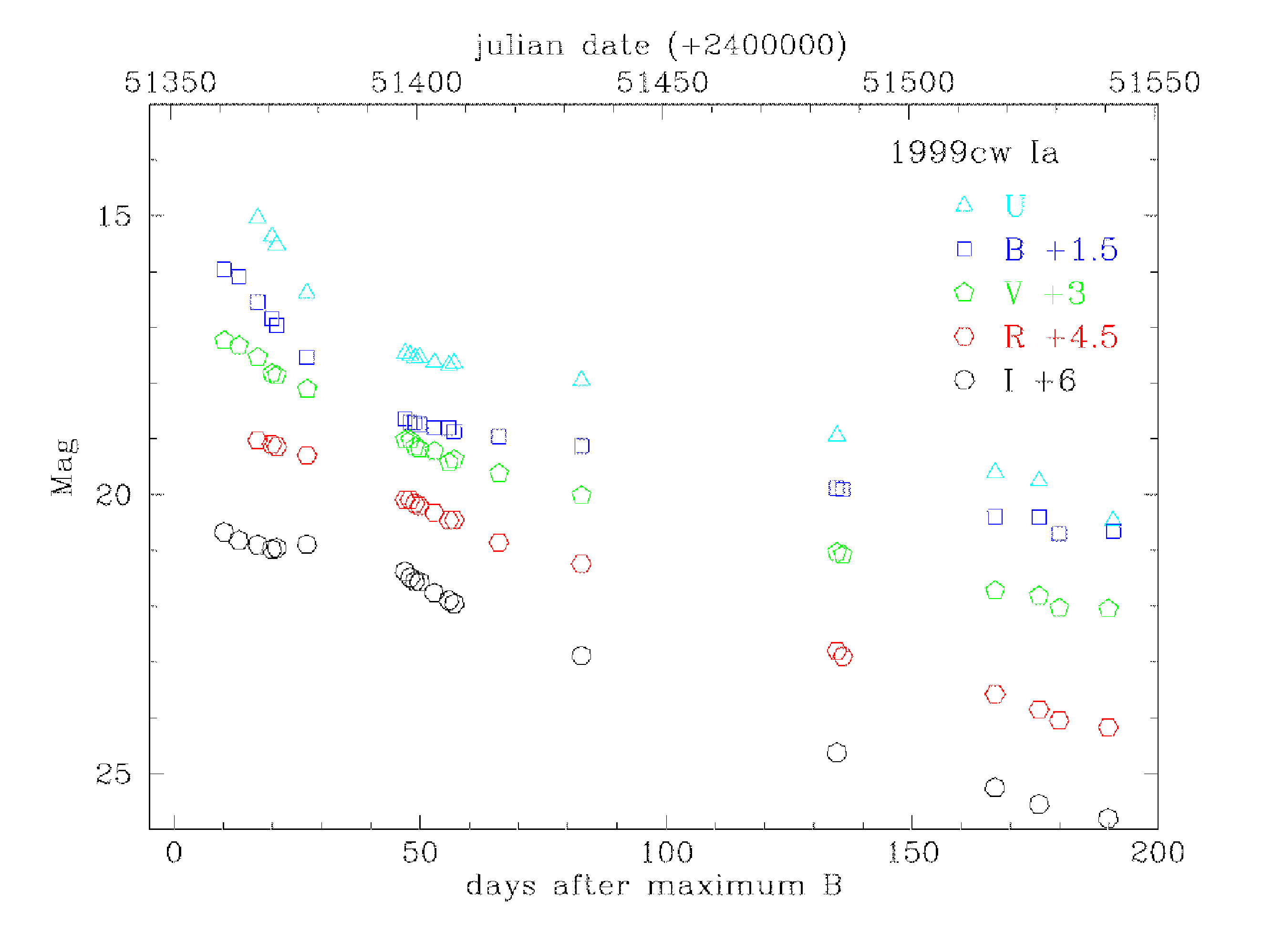}
\includegraphics [width=5cm, height=6.6cm,angle=270]{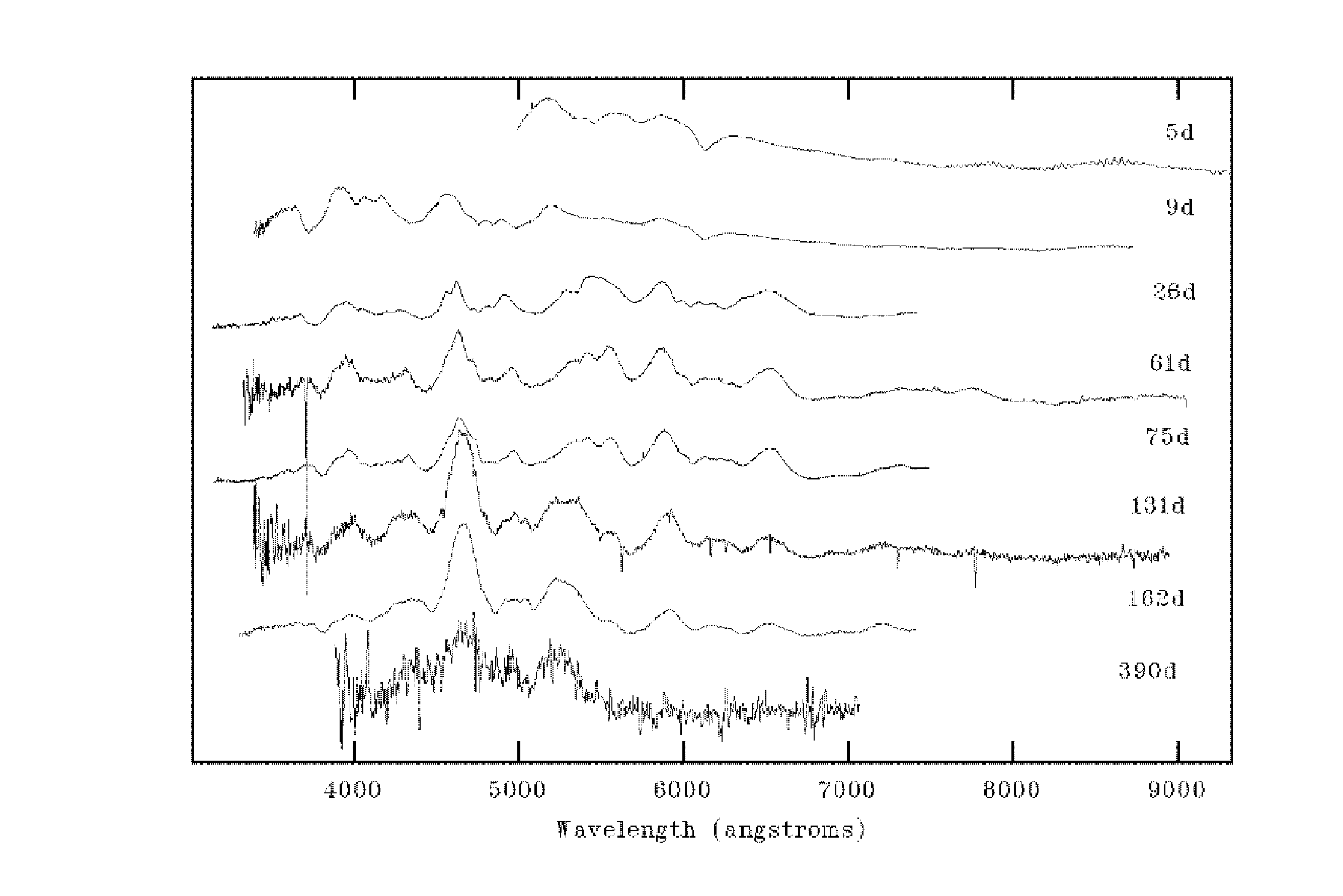}
\caption{Left: UBVRI light curves of SN1999cw.The magnitudes for each band are shifted by
the amounts shown in the plot. Right: evolution of the optical spectra corrected for redshift and reddening.   }\label{figura}
\end{center}
\end{figure}

Photometric data have been reduced by using PSF technique by mean of the SNOoPY procedures (Patat, 1995),
 which are based on IRAF tasks. The results have been checked and compared with
 different photometric reduction methods, namely Daophot and Template Subtraction producing
 comparable results(Bufano, 2004). The resulting UBVRI light curves are reported in Fig \ref{figura}.
 SN1999cw is well sampled, but it is clear that observations began after
 the epoch of maximum light, even if they cover a very long period of 
the supernova evolution. Light curves are characterized by a relatively 
slow luminosity decline rate and a secondary maximum light in R and I bands. 
In order to define the epoch of maximum, we have applied the method by
 Altavilla et al (2004), which minimize the $\chi^{2}$ value with respect to a number
 of templates. The best fit is obtained with slow declining SNe Ia, similar 
to SN1991T. In particular, SN1991T and SN1999cw have 
same shape and post-maximum decline rate $(\Delta m_{15}(B)=0.94)$. By overplotting 
the light curves, we find that the B light curve maximum occured on $JD=2451355.2 (\pm 0.5 )$
  $(25^{th} June 1999)$ at $  B_{max}  = 14.30 (\pm 0.05) mag.$ Thus our observations began five days after maximum light.
 By fitting V and R light curves, we obtain
               $V_{Max}= 14.25 (\pm 0.05) mag $ and $
               R_{Max}= 14.12 (\pm 0.05)mag.$\\
To derive the reddening of SN1999cw we used $B-V$ color curve and the  
empirical relation of Phillips (1999). We found that SN1999cw did
 not suffer of any significant dust absorption inside to the host galaxy. 
 So the absolute B magnitude is
$M_{B_{Max}} = B_{Max} - \mu - A_{B}^{g} = -19.41 (\pm 0.13) mag$
where $\mu= 33.56  (H_{0}=71 km/s·Mpc)$ and $A_{B}^{g} = 0.154$. This value is in good 
agreement with the value, $-19.69 (\pm0.31)$ mag, computed by using Altavilla's 
empirical relation $M_{B}$ vs $\Delta m_{15}(B)$, based on nearby SNe for which distance 
can be determinate by Cepheids.\\

The spectral evolution of SN1999cw is plotted in Fig 1. Because of the lack
 of information around the maximum, we cannot study the appearance of  SiII
 lines (or other intermediate-mass element lines) typical of the SNe
 1991T-like around this phase. The SiII $(\lambda 6355)$ absorption line is already present in the
 first spectrum, but it is particularly weak. By using our new procedure
 "PASSparToo" (Harutyunyan, this conference), we compared SN1999cw spectra
 with all the data in the Padova-Asiago Supernovae archive and verified that the best fit
 is with over-luminous Type Ia SNe.  By measuring the minimum of the SiII
 absorption line in the first spectrum, we deduce an expansion velocity
 v$\simeq$10.700 km/s. In the following spectra, the velocity remained nearly
 costant, in analogy to the velocity evolution of SN1991T and similar objects (Benetti, this conference).

In conclusion, the preliminary analysis of the optical data of SN1999cw shows that this 
object has photometric and spectroscopic behavior very similar to SN1991T.

\vspace{0.15cm}
\textbf{\small{References}}\\
\small{ Altavilla, G.; Fiorentino, G.; Marconi, M. et al  2004 MNRAS 349,1344}\\
\small{Bufano, F. 2004, Tesi di Laurea }\\
\small{Johnson, R. \& Li, W.D. 1999 IAUC.7211}\\
\small{ Patat, F. 1995, Tesi di Dottorato }\\
\small{Phillips, M.M.; Lira, P.; Suntzeff, N. et al 1999 AJ 118,1766}\\
\small{Rizzi, L.; Patat, F.; Benetti, S.; Cappellaro, E.; Turatto, M. 1999 IAUC.7216} \\
\small{ Schlegel, D.J.; Finkbeiner, D.P.; Davis, M. 1998 ApJ 500,525}\\
\end{document}